# Cross-Shaped Windows Transformer with Self-supervised Pretraining for Clinically Significant Prostate Cancer Detection in Bi-parametric MRI

Yuheng Li[1,2], Jacob Wynne[1], Jing Wang[1], Richard L.J. Qiu[1], Justin Roper[1], Shaoyan Pan[1], Ashesh B. Jani[1], Tian Liu[3], Pretesh R. Patel[1], Hui Mao[2,4] and Xiaofeng Yang[1,2*]

[1]Department of Radiation Oncology and Winship Cancer Institute, Emory University, Atlanta, GA

[2]The Wallace H. Coulter Department of Biomedical Engineering, Emory University and Georgia Institute of Technology, Atlanta, GA

[3]Department of Radiation Oncology, Icahn School of Medicine at Mount Sinai, New York, NY

[4]Department of Radiology and Imaging Science and Winship Cancer Institute, Emory University, Atlanta, GA

*Corresponding to: xiaofeng.yang@emory.edu

**Abstract**

Bi-parametric magnetic resonance imaging (bpMRI) has demonstrated promising results in prostate cancer (PCa) detection using convolutional neural networks (CNNs). Recently, transformers have achieved competitive performance compared to CNNs in computer vision. Large-scale transformers need abundant annotated data for training, which are difficult to obtain in medical imaging. Self-supervised learning (SSL) utilizes unlabeled data to generate meaningful semantic representations without the need for costly annotations, enhancing models' performance on tasks with limited labeled data. We introduce a novel end-to-end Cross-Shaped windows (CSwin) transformer UNet model, CSwin UNet, to detect clinically significant prostate cancer (csPCa) in prostate bi-parametric MR imaging (bpMRI) and demonstrate the effectiveness of our proposed self-supervised pre-training framework. Using a large prostate bpMRI dataset with 1500 patients, we first pretrain CSwin transformer using multi-task self-supervised learning to improve data-efficiency and network generalizability. We then finetune using lesion annotations to perform csPCa detection. Five-fold cross validation shows that self-supervised CSwin UNet achieves $0.888\pm0.010$ AUC and $0.545\pm0.060$ Average Precision (AP), significantly outperforming four comparable models (Swin UNETR, DynUNet, Attention UNet, UNet). Using a separate bpMRI dataset with 158 patients, we evaluate our method's robustness to external hold-out data. Self-supervised CSwin UNet achieves 0.79 AUC and 0.45 AP, still outperforming all other comparable methods and demonstrating good generalization to external data.

## 1. INTRODUCTION

Prostate cancer (PCa) is the most common non-cutaneous cancer in men[1]. PCa treatments, including surgery, radiotherapy, and anti-androgen therapy, are heavily influenced by the histologic grade[2]. As high-grade, clinically-significant PCa often leads to treatment failure, accurate PCa grading is vital for effective treatment planning[3].

Multiparametric MRI (mpMRI), which combinines anatomical and functional MR imaging sequences, has become an important tool for noninvasive PCa detection [4-6]. Current prostate MRI interpretation is based on Prostate Imaging Reporting and Data System: Version 2 (PI-RADS v2), which still remains qualitative or semi-quantitative and does not eliminate the possibility of inter-reader variability [7]. csPCa can appear as lesions of various shapes and sizes, often resembling benign conditions. Without skilled radiologists, these similarities can result in poor agreement among readers and inadequate interpretations. Consequently, developing accurate and robust algorithms for detecting csPCa has become an important task within medical image analysis. Several studies have proposed radiomics models or deep learning models to automatically detect and classify csPCa [8].

Deep learning approaches have demonstrated remarkable success in medical image analysis, with vision transformer showing competitive performance compared to convolutional neural networks [9,10]. Local-windowed transformers [11,12] have been introduced to improve computational efficiency, but they only enlarge the receptive field slowly and may hinder its potential in dense segmentation tasks. The cross-shaped window transformer achieves strong global dependency by performing self-attention across horizontal and vertical stripes, significantly extending the receptive field with minimal computational expense[13]. It segments the input into equal-width stripes, adjusting the width according to the network's depth to optimize performance.

However, transformer-based models in medical imaging still face challenges due to the need for large labelled datasets, which are costly in terms of time and expenses [14]. Without sufficient labelled data for supervised training, deep learning models tend to have low generalizability across different clinical settings, causing a significant loss of model performance. As an alternative, self-supervised learning aims at learning useful representations from unlabeled data [15-17]. Recent studies have demonstrated that self-supervised pretraining of transformer models can significantly enhance performance for downstream tasks [18] and improving network generalization ability[19]. Previous SSL methods such as Tang et al. [20] simply combined a series of pretext tasks to learn semantic representations from CT images, which neglects the intrinsic relationships among each pretext task. Intuitively, if the model can learn a shared embedding among different pretext tasks, this parameter can automatically adjust the difficulty of each pretext task to facilitate learning better representations. In this study, we introduce a multi-task learning loss that automatically adjusts the importance of each pretext task, enhancing the representations learned for detecting csPCa. This method also simplifies the time-consuming process of determining the optimal weight for each pretext task's hyperparameters.

In this paper, we propose an end-to-end 3D transformer model to detect csPCa in prostate bpMRI and a multi-task learning self-supervised pretraining framework. The overall workflow of our method is shown in Fig.1. The contributions of this work include:

- We develop a novel architecture CSwin UNet. Our CSwin encoder performs self-attention in horizontal, vertical and longitudinal stripes. By splitting self-attention in three independent dimensions. We also introduce scaled cosine attention into our transformer backbone, which improves the detection metrics. Our architecture outperforms comparable CNN and transformers in 3D csPCa detection.

- We propose a multi-task learning loss to unify the pretraining pipeline with self-supervised pretext tasks including contrastive learning [21], context restoration [22], and rotation prediction [23]. We study the effects of self-supervised pretraining on the downstream csPCa detection task by analyzing the data efficiency using limited training data.
- We validate the performance of Cswin UNet on a large public bpMRI dataset PI-CAI containing 1500 scans from 1476 patients across three institutions in The Netherlands [24]. Each patient underwent histopathology and had follow-up ($\geq$ 3 years) as the reference standard. Our model achieves state-of-the-art among both CNN methods and transformer methods.
- Using Prostate158 [25] as external data, we demonstrate that our self-supervised pretraining strategy yields more powerful semantic representations and improves model robustness compared to fully-supervised method.

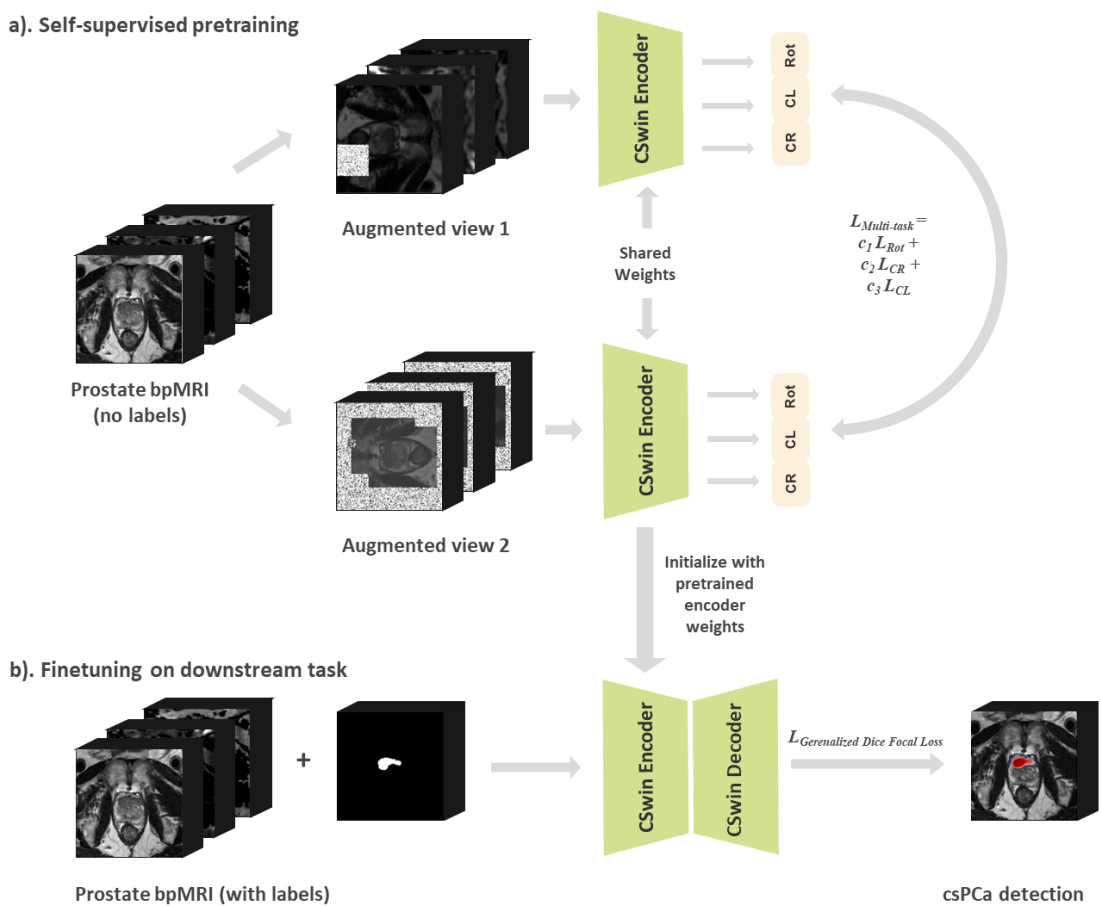

**Figure 1.** Overview of our proposed (a) self-supervised pretraining and (b) finetuning workflow for csPCa detection. Each unlabelled prostate bpMRI image underwent data augmentations twice to generate two separate views with similar semantics information. Then, CSwin encoder is pretrained on unlabeled data using three pretext tasks: contrastive learning (CL), context restoration (CR), and rotation prediction (Rot) with automatic weighted loss. Finally, we fine-tune the model with labelled bpMRI to csPCa detection using pre-trained CSwin encoder.

## 2. METHODS

## 2.1 CSwin UNet

CSwin UNet comprises a CSwin Transformer encoder that directly utilizes 3D patches and is connected to a CNN-based decoder via skip connections at different resolutions. Fig. 2 illustrates the overall architecture of CSwin UNet. We describe the details of the encoder and decoder in this section.

### 2.1.1 CSwin Encoder

Given input volume $X \in \mathbb{R}^{H \times W \times D \times C}$, where H, W, D are the height, width, and depth, respectively and C represents the image dimension, we leverage the overlapped convolutional token embedding to obtain patch tokens $T \in \mathbb{R}^{\frac{H}{2} \times \frac{W}{2} \times \frac{D}{2} \times F}$ with dimension F. Next, an additional convolutional token embedding layer is used to further reduce the patch tokens $T$ to size of $\frac{H}{4} \times \frac{W}{4} \times \frac{D}{4}$. To produce a hierarchical representation, the whole network consists of 4 stages. At the start of each stage, a convolution layer (kernel 3, stride 2) is used to reduce the number of tokens and double the channel dimension. For stage $i$, the constructed feature maps have $\frac{H}{2^{i+1}} \times \frac{W}{2^{i+1}} \times \frac{D}{2^{i+1}}$ tokens. CSwin Transformer Block has an overall topology similar to the vanilla multi-head self-attention Transformer block with two important differences: 1) It replaces the self-attention mechanism with Cross-Shaped Window Self-Attention; 2) A scaled cosine attention function is used to replace the previous dot-product attention.

We formulate the cross-shaped window attention in 3D by splitting 3D volumes into horizontal, vertical, and longitudinal stripes (Fig.3). Formally, the input token T will be first linearly projected to G heads, which are then equally split into three parallel groups (each has $\frac{G}{3}$ heads). The first group of heads perform horizontal self-attention, the second group performs vertical self-attention and the third group performs longitudinal self-attention. Finally, the output of these three parallel groups will be concatenated. To calculate horizontal self-attention, $T$ is evenly partitioned into non-overlapping horizontal stripes of equal width $sw$ with $sw \times W \times D$ windows. Similarly, the T can also be independently partitioned into non-overlapping vertical and longitudinal windows. We explicitly defined the horizontal, vertical, longitudinal windows as $P_h, P_v, P_l$, respectively. The stripe width $sw$ can be adjusted to balance the learning capacity and computational complexity. Formally, suppose the projected queries ($Q$), keys ($K$) and values ($V$) of the $k^{th}$ head all have dimension $d_k$, then the output of the horizontal, vertical and longitudinal stripes self-attention for $k^{th}$ head is defined as:

$$T = [P_h^1, \ldots, P_h^M, P_v^1, \ldots, P_v^M, P_l^1, \ldots, P_l^M], \quad (1)$$

$$A_k^i = Attention(P^i W_k^Q, P^i W_k^K, P^i W_k^V), \quad (2)$$

$$H-Attention_k(X) = [P_k^1, P_k^2, \ldots, P_k^M], \; for \; k = 1,2,\ldots,\frac{G}{3}, \quad (3)$$

$$V-Attention_k(X) = [P_k^1, P_k^2, \ldots, P_k^M], \; for \; k = \frac{G}{3}+1,\ldots,\frac{2G}{3}, \quad (4)$$

$$L-Attention_k(X) = [P_k^1, P_k^2, \ldots, P_k^M], \; for \; k = \frac{2G}{3}+1,\ldots,G, \quad (5)$$

where $P^i \in \mathbb{R}^{(sw \times W \times D) \times F}$ and $M = \frac{H}{sw}, i = 1,2,\ldots,M$. $W_k^Q \in R^{F \times d_k}, W_k^K \in R^{F \times d_k}, W_k^V \in R^{F \times d_k}$ represent the projection matrices of $Q, K, V$ for the $k^{th}$ head respectively and $d_k$ is set as $\frac{C}{k}$ with output dimension $C$. The final CSwin-Attention is calculated as:

$$CSwin-Attention(X) = Concat(H-attention_1, \ldots, V-attention_1, \ldots L-attention_1, \ldots)W^o$$

$W^o \in R^{C \times C \times C}$ is the commonly used projection matrix that projects the self-attention results into the target output dimension. The attention area of each token within one transformer block is enlarged via multi-head grouping, while existing self-attention mechanisms apply the same self-attention operations across different multi-heads.

We propose to integrate scaled cosine attention into our transformer blocks. When calculating original self-attention, similarity terms of the voxel pairs are computed as a dot product of the query and key vectors. However, in the norm res-post-norm configuration, this approach could lead to the learned attention maps of some blocks and heads being frequently dominated by a few pixel pairs. As a solution, a scaled cosine attention is proposed that computes the attention logit of a voxel pair $i$ and $j$ by a scaled cosine function:

$$Sim(q_i, k_j) = \frac{cos(q_i, k_j)}{\tau} + B_{ij}, \quad (6)$$

where $B_{ij}$ is the positional bias between $i$ and $j$ and $\tau$ is a learnable scalar. $q_i, k_j$ are the query and key vectors obtained by a linear transformation of the input. The cosine function is normalized so that the attention values are less likely to be extreme.

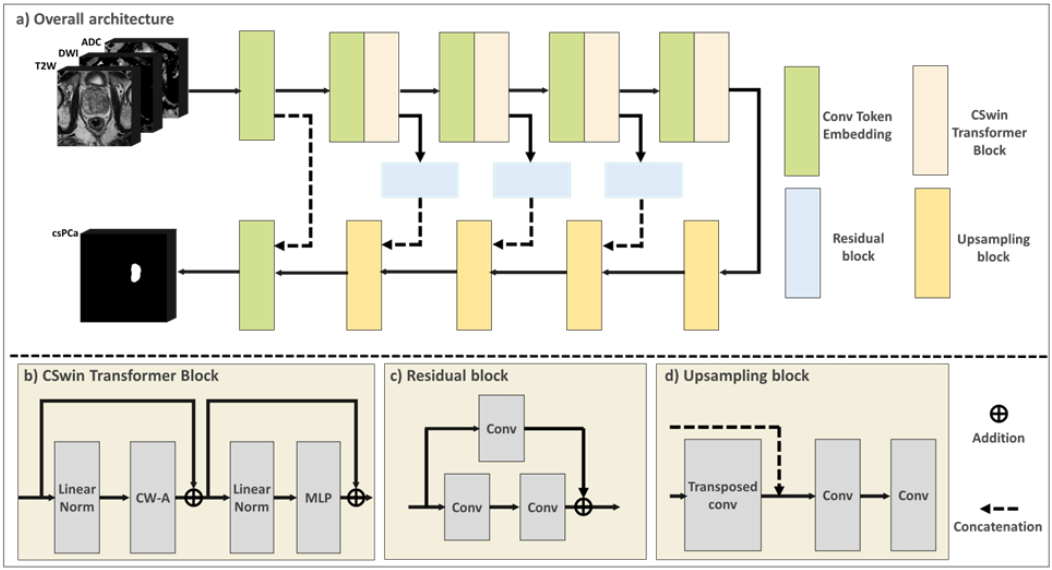

**Figure 2.** Proposed structure of 3D CSwin UNet for detecting csPCa. a) The overall network consists of a CSwin encoder and a CNN decoder. CSwin encoder sequentially down-samples the input by a ratio of 1/4, 1/8, 1/16, and 1/32. The decoder mirrors the configurations of the encoder and up-samples the features. The last stage Conv Token Embedding maintains the spatial resolution of the feature.

### 2.1.2 Decoder

The decoder is a CNN-based module connected to the encoder at each stage via skip connections. At each stage $i \in [1,2,3,4,5]$, the output sequence of representations in the encoder and the bottleneck are reshaped into features of size $\frac{H}{2^i} \times \frac{W}{2^i} \times \frac{D}{2^i}$. The extracted representations at each stage are then fed into a residual block consisting of two $3 \times 3 \times 3$ convolutional layers with instance normalization. The processed features from each stage are then upsampled by using a deconvolutional layer and concatenated with processed features of the preceding stage. The concatenated features are fed into a similar residual block. For segmentation, we concatenate the output of the CSwin encoder with processed features of the input volume and feed them into a residual block followed by a final $1 \times 1 \times 1$ convolutional layer with a proper activation function (i.e. softmax) for computing segmentation probability maps.

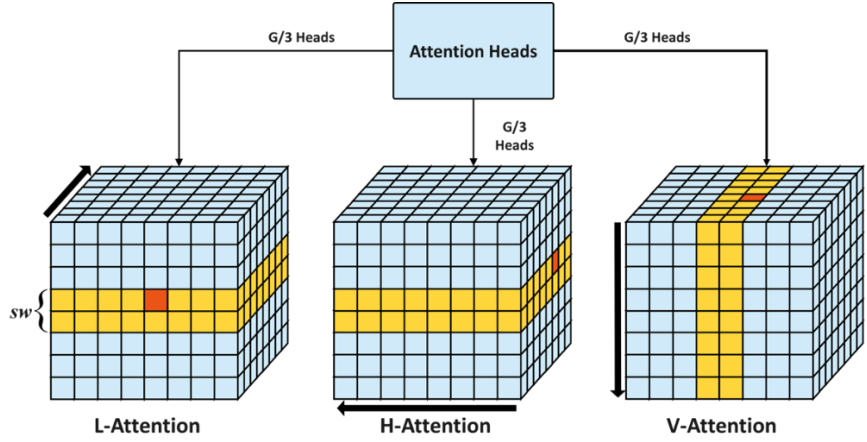

**Figure 3.** Cross-shaped window attention mechanism for 3D data. First, we split *G* multi-heads into three groups and calculate self-attention in longitudinal, vertical and horizontal directions simultaneously (*L-, V-* and *H-Attention*). Second, we adjust the stripe width *sw* according to the depth network, which can achieve better trade-off between computation cost and capability.

### 2.2 Self-supervised learning with multi-task learning loss

In this section, we discuss our self-supervised learning strategy from bpMRI scans, requiring no manual annotations in the pretraining phase. Inspired by Tang et al. [20], we chose three pretext tasks that achieved promising results: contrastive learning [21], context restoration [22] and rotation prediction [23].

#### 2.2.1 Contrastive learning

Contrastive learning aims to maximize mutual information between semantically similar (positive) samples and minimize that between non-similar (negative) samples in a shared latent space [26,27]. Contrastive encoding is obtained by attaching a linear layer to the CSwin UNet encoder, which maps each augmented sub-volume to a latent representation. We formulate contrastive loss as described in [21] with cosine similarity as the distance measurement of encoding representation. Formally, we define contrastive loss $L_{CL}$ for a pair of embedding vectors $v_i, v_j$ as:

$$L_{CL} = -log \frac{exp\left(\frac{sim(v_i, v_j)}{t}\right)}{\sum_{k}^{2N} 1_{k \neq i} \, exp\left(\frac{sim(v_i, v_j)}{t}\right)}$$

where *t* is the measurement of normalized temperature scale. 1 is the indicator function evaluating to 1 if and only if $k \neq i$. *sim* denotes the dot product between normalized embeddings.

#### 2.2.2 Context restoration

Context restoration aims to restore missing patches from the corrupted image to learn semantic information [22]. In our work, we consider a combination of patch shuffling and cut-out augmentations for 3D prostate bpMRI data. We mask a sub-volume $X \in R^{H \times W \times D \times F}$ randomly with volume ratio *s*. Then, we select *t* voxels within *X* and shuffle their context. To restore context in the images, we attach a transpose convolution layer to the encoder as the reconstruction head and denote its output as $\hat{X}^M$. The reconstruction objective is defined by an L1 loss between *X* and $\hat{X}^M$:

$$L_{CR} = \|X - \hat{X}^M\|_1$$

### 2.2.3 Rotation prediction

Originally proposed by Gidaris *et al.* [23], the rotation prediction task encourages the model to learn visual representations by simply predicting the angle by which the input image is rotated. The intuition behind this task is that for a model to successfully predict the angle of rotation, it needs to learn enough semantic information about the image. We consider the range of angles to be multiples of 90 degrees (0°, 90°, 180°, 270°, along the z-axis of the 3D coordinate system $(x, y, z)$, so a total of 4 possible rotations is predicted. Formally, we minimize the cross-entropy loss $L_{rot}$:

$$L_{rot} = -\sum_{r=1}^{R} y^k \log(\hat{y}^k)$$

where k ∈ [1, .., K] is an arbitrary rotated 3D image from the list of K rotated images and $(y^k, \hat{y}^k)$ are the true rotation angle and the predicted rotation angle respectively.

### 2.2.4 Multi-task learning using automatic weighted loss

In a common encoder-decoder architecture, multi-task learning aims to find a common representation in the encoder stage of the network, while the individual tasks τ ∈ T are solved in their respective decoder branches of the network. However, previous approaches [20] use a naïve weighted sum of losses where the weights for each loss are manually tuned, requiring significant computational resources and time. We combine the loss function of each pretext task using the concept of auxiliary tasks [28]. By choosing auxiliary tasks that help the network learn a rich and robust common representation of the image, we can boost its performance on the main task of prostate cancer segmentation. Formally, to optimize a multi-task network for the learnable parameters $\omega_T$, we define multi-task learning loss function $L_{mult}$ with learnable weight coefficients $c_{1,2,3}$:

$$L_{task}(x, y_T, \hat{y}_T, \omega_T) = \sum_{t \in T} c_t L_t(x, y_t, \hat{y}_t, \omega_t)$$

where *t* represents each individual task, is the weight for each task, $y_t$ is the ground truth, $\hat{y}_t$ is the predicted label. Instead of manually tuning $c_t$ to find the optimal weights, the coefficient can be added to the learnable network parameters $\omega_T$. A regularization term $\ln(1 + c_T^2)$ was added to prevent trivial solution and enforce positive regularization values. Finally, we reformulate the multi-task learning loss as:

$$L_{task}(x, y_T, \hat{y}_T, \omega_T) = \frac{1}{2c_1^2} L_{CL} + \frac{1}{2c_2^2} L_{CR} + \frac{1}{2c_3^2} L_{rot} + \ln(1 + c_1^2)(1 + c_2^2)(1 + c_3^2)$$

During pretraining, three projection heads tailored for each pretext task are attached to the CSwin encoder which is trained with our proposed multi-task learning loss. We use a composition of data augmentation operations to yield effective representation, including random rotation, random cutout, pixel shuffling, contrast adjustment and bias field estimation. During pretraining, we apply data augmentation twice to the input data generating two different views of the input for contrastive learning. An illustration of these data augmentation techniques is shown in Figure 4.

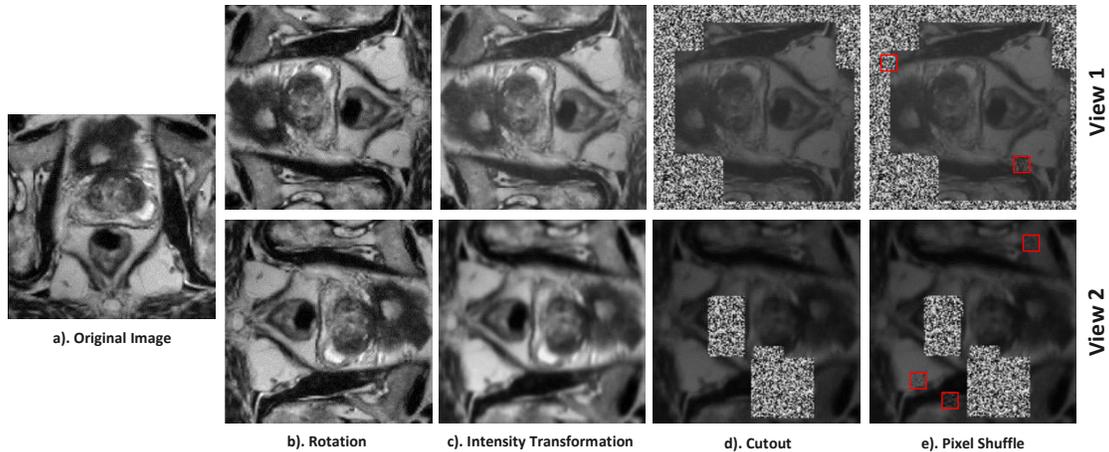

**Figure 4.** Illustrations of data augmentation for self-supervised learning. (a) shows the original T2W MRI image. (b) shows image rotated by 90 degrees along z-axis. (c) shows result of intensity transformation (gaussian blur, contrast adjustment). (d) shows the random cutout regions to be restored. (e) shows the result of pixel shuffling (red rectangles indicate shuffled pixels).

### 2.4 Implementation details

We train CSwin UNet using AdamW optimizer with a cosine annealing rate scheduler. For self-supervised pretraining, the network is trained for 300 epochs with an initial learning rate of $10^{-3}$ and a linear warmup of 20 epochs. We use a batch size of 16 because contrastive learning requires a large batch size to mine negative samples. Pretraining experiments are performed using a single NVIDIA A100 GPU.

For csPCa detection, we initialize the CSwin encoder with the weights from self-supervised pretraining. For loss function, we use generalized dice focal loss to address the class imbalance at voxel level. We compute a weighted average of generalized dice loss and focal loss. The network was trained for 150 epochs with an initial learning rate of $10^{-4}$ and a linear warmup of 10 epochs. Hyperparameters are tuned using the first fold. Training is conducted on a single NVIDIA Quadro RTX 5000 GPU. Our model is implemented in PyTorch with MONAI.

For pre-training tasks, we adopt the following parameters: 1). Contrastive learning: a feature size of 384 is used as the embedding dimension; 2). Context restoration: the cut-out ROI vs image ratio is randomly chosen from 0.1-0.48 for every iteration. Next, 14 patches of (12, 12, 4) within the image were selected to shuffle voxels within every region; 3). Rotation prediction: the rotation degree is 0°, 90°, 180°, 270° along the z-axis of the 3D coordinate system $(x, y, z)$.

### 3. Experiments and results

### 3.1 Dataset and pre-processing

In this study, two datasets with bpMRI scans (axial T2-weighted (T2W), high b-value ($\geq$ 1400) diffusion-weighted imaging (DWI) and apparent diffusion coefficient (ADC) maps) for prostate cancer detection were used.

### 3.1.1 PI-CAI dataset

To train and validate our model, we used the PI-CAI (Prostate Imaging: Cancer AI) challenge training set with 1500 anonymized prostate bpMRI scans from 1476 patients, acquired between 2012-2021, at three centers (Radboud University Medical Center, University Medical Center Groningen, Ziekenhuis Groep

Twente) based in The Netherlands. Patient cases were annotated with histologically-confirmed findings: Gleason grade group ⩾ 2 as positives, and Gleason grade group ⩽ 1 or PI-RADS ⩽ 2 as negatives. For all cases, csPCa lesions were delineated by one of 10 trained investigators or 1 radiology resident.

Out of the 1500 cases, 1075 cases have benign tissue or indolent PCa and 425 cases have csPCa. Out of these 425 positive cases, only 220 cases carry an annotation derived by a human expert. The remaining 205 positive cases have not been annotated. We first divide the labelled data (1295 cases) into 5 folds, following the challenge organizers. During self-supervised pretraining, we perform pretraining on 4 folds and the 205 unlabelled cases. During fully-supervised training, we only finetune our model using the 4 folds with labels, while internally testing the model's performance on the last 1 fold. We repeat the pretraining and finetuning process 5 times serving as cross validation.

### 3.1.2 Prostate158 dataset

In addition, we used a second public dataset (Prostate158) as external validation. Prostate158 consists of 158 expert-annotated 3T prostate MRIs comprising T2w sequences and DWI sequences with ADC maps similar to PI-CAI. All patients were examined at a German university hospital (Charit´e University Hospital Berlin) between February 2016 and January 2020. MR images were acquired on Siemens VIDA and Skyra clinical 3T scanners (Siemens Healthineers, Erlangen, Germany) according to an acquisition protocol that complies with current guidelines and using B1 shimming. T2w sequences were acquired with slice thickness 3 mm, no interslice gap, and in-plane resolution $0.47 \times 0.47$ mm. For DWI, the acquisition parameters were as follows: slice thickness 3 mm, no interslice gap, in-plane resolution $1.4 \times 1.4$ mm.

Two board-certified radiologists with 6 and 8 years of experience in uro-oncologic imaging annotated all MR images. Pixel-wise segmentations were provided for the central gland (central zone and transitional zone), peripheral zone, and PCa lesions, which were defined as suspicious areas with a PIRADS score ≥ 4. All PCa lesions were segmented in the ADC map and correlated with T2w sequences and DWI high b-value images.

### 3.1.3 Pre-processing

We spatially resample all bpMRI scans to a common axial in-plane resolution of 0.5 mm$^2$ and slice thickness of 3.6 mm via B-spline interpolation. We then perform a center crop of the lesion, with a size of 72.0 mm $\times$ 72.0 mm $\times$ 57.6 mm or $144 \times 144 \times 16$ voxels. To fit the input requirement of our network, all images are resized to $160 \times 160 \times 32$. We perform z-score normalization for each T2W and DWI scans and global z-score normalization for ADC scans.

## 3.2 Experiments

### 3.2.1 Performance analysis

Following previous works in PCa detection [29-31], we evaluate patient-based diagnosis using Receiver Operating Characteristic (ROC) curve, summarized to the area under the ROC curve (AUROC). For lesion-level diagnosis, we generate Precision-Recall (PR) curve, summarized to Average Precision (AP) by calculating the area under curve.

To obtain detection maps, we use automatic postprocessing tools developed by Bosma et al. [32] on model segmentation outputs. This process generates a lesion candidate by selecting the highest confidence predictions from the model's softmax output. It then includes all 3D connected components that have at least 40% of the peak confidence level. After selecting a lesion candidate, it is removed from the softmax maps. This procedure repeats for all potential lesion candidates until there are no remaining softmax predictions. True positive is defined as lesions sharing a minimum of 0.1 dice score with ground-truth annotations following [29,33,34], since most csPCa lesions are small with indistinct margins and have large inter-reader variability in their interpretation. If a detected lesion did not intersect ground truth lesion, it is considered as a false positive.

We estimate confidence intervals as twice the standard deviation from the mean of 5-fold cross-validation (applicable to validation sets). We verified statistically significant improvement with a p-value using a Wilcoxon signed-rank test on the difference in patient-level AUROC and lesion-level AP [32]. We apply the Holm-Bonferroni method to adjust the p-value and set the threshold for statistical significance at 0.005 [35].

For Prostate158, we selected the best performing model in the validation set to evaluate on the entire Prostate158 dataset. For statistical tests, we conducted paired-t tests in patient-level AUROC and lesion-level pAUC, also with threshold for statistical significance at 0.005.

### 3.2.2 Performance on PI-CAI

We first evaluated our proposed methods on PI-CAI validation sets. For comparable methods, we report results from 3 other state-of-the-art CNN segmentation models and 1 state-of-the-art transformer model. The first one is the basic UNet [36]. The second model is Attention UNet [37], a U-Net model with attention gates to suppress irrelevant regions while highlighting salient features. The third model is DynUNet implemented in MONAI, which followed nnUNet architecture [38]. The transformer model we compare to is Swin UNETR [20] which has a Swin transformer encoder and CNN decoder. We facilitated a fair comparison by maintaining an identical preprocessing, augmentation, tuning and train-validation pipeline for each candidate system in a given experiment.

**Ablation on network components**: First, we ablate on the effectiveness of cosine attention in CSwin UNet. Cosine similarity has shown to be an effective normalization method [39] compared to dot-product attention. As shown in Table 1, cosine attention improved csPCa detection by 1.3% in AUC and 0.5% in AP. Second, dynamic stripe width also affects the performance of CSwin encoder. As shown in Table 1, we find that increasing stripe width also improves network performance.

**Table 1.** Comparison of different CSwin architectures. We study the effect of cosine attention and dynamic stripe width (*sw*) on csPCa detection.

| Model | AUC | AP |
|---|---|---|
| No cosine attention (fix sw = 2) | 0.861 ± 0.010 | 0.516 ± 0.049 |
| Cosine attention (fix sw = 2) | 0.874 ± 0.014 | 0.521 ± 0.045 |
| **Cosine attention (sw = 1,2,5,5)** | **0.880 ± 0.010** | **0.543 ± 0.042** |

**Patient-based diagnosis:** From ROC analysis on PI-CAI dataset (Fig.5[40]), our proposed CSwin UNet reached 0.880±0.013 AUC and our self-supervised CSwin UNet reached 0.888±0.010 AUC in patient-level diagnosis, ahead of all other candidate systems by a margin of 4.6–6.1%. Without self-supervised pretraining, our CSwin UNet performed significantly better than U-Net ($p<0.005$), Attention UNet ($p<0.005$), Dyn-UNet ($p<0.005$) and Swin UNETR ($p<0.005$). After self-supervised learning, our model still performed significantly better than all comparable methods ($p<0.005$). Our proposed methods achieved comparable performance to the detection system developed by [29] which achieved 0.882±0.03 AUC.

**Lesion-based localization:** From PR analysis on PI-CAI dataset (Fig.6[40]), our CSwin UNet model achieved 0.543±0.042 AP, significantly outperforming UNet ($p<0.005$), Attention UNet ($p<0.005$), Dyn-UNet ($p<0.005$) and Swin UNETR ($p<0.005$). After self-supervised pretraining, our model reached 0.545±0.06 AP, still significantly outperforming comparable methods ($p<0.005$). Visualization of csPCa detection maps (Fig.7) further confirms the superior performance of CSwin UNet over other networks. Self-supervised pretraining improved CSwin's ability to detect small lesions.

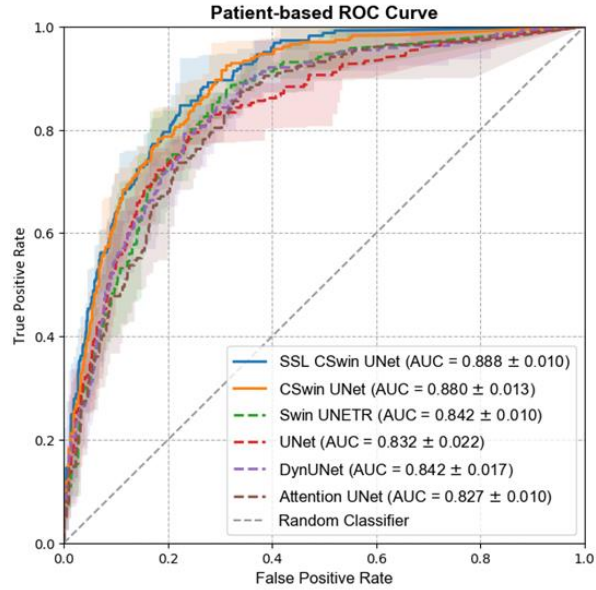

**Figure 5**. Patient-based ROC analysis of csPCa detection in PI-CAI using the proposed models and other comparable methods. Transparent areas indicate the 95% confidence intervals.

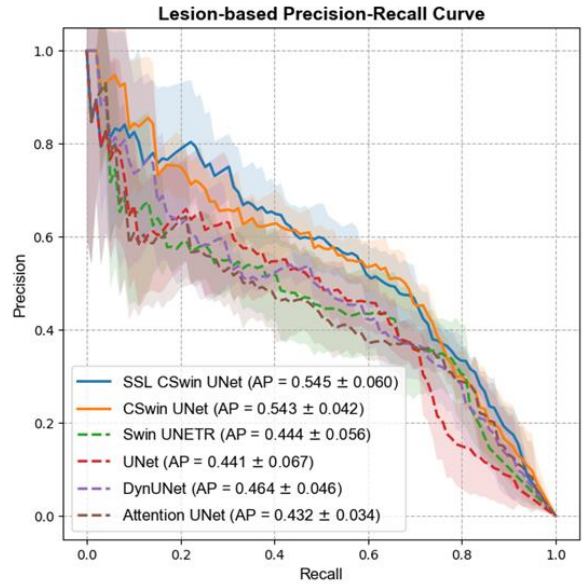

**Figure 6.** Lesion-based PR analysis of csPCa detection in PI-CAI using the proposed models and other comparable methods. Transparent areas indicate the 95% confidence intervals.

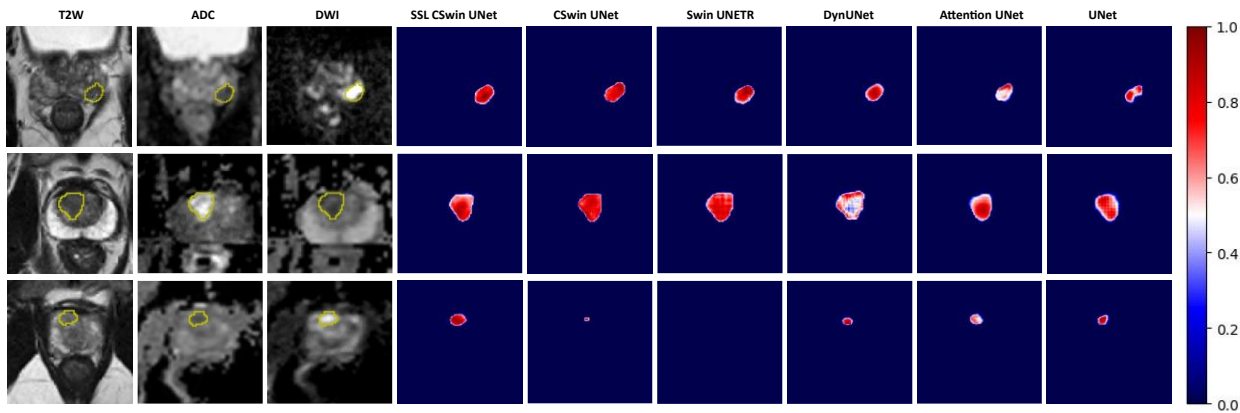

**Figure 7**. Visualization of csPCa detection maps from our proposed systems and comparable methods. Gradient-weighted class activation maps (GradCAM) and their corresponding T2W, DWI and ADC scans for three patient cases from PI-CAI validation set are shown above. 3D GradCAMs were generated from csPCa segmentation maps and activation levels were normalized to (0,1). CSwin UNet detected csPCa with clear boundary and good overlap. Self-supervised learning improved detection of small lesions (***bottom row***).

### 3.2.3 Sample efficiency of self-supervised finetuning

We assess the effectiveness of SSL by quantifying gains in performance metrics under 1). different pretext tasks and 2). at different sample sizes. For our self-supervised pretraining, we experiment with different combinations of each pretext task. Then, we finetune our proposed CSwin model by randomly selecting subsets of patients at 25%, 50%, and 100% from the training set described in Section 3.1.1. Training using random initialization is used as baseline in each comparison group. We also conduct ablation experiments on our proposed automatic weighted loss. Table 2 details the performance of CSwin UNet under different configurations.

**SSL pretraining improves sample efficiency.** With pretraining, our model outperforms baselines with randomly initialized weights. We note that the gains in performance increase when using less labeled training data. At 25% labeled data, our SSL pretraining method increased AUC by 4.79% and AP by 4.25%. At 50% labeled data, our proposed method increased AUC by 2.68% and AP by 3.36%. However, when using 100% labeled data for finetuning, the gains in model performance become relatively negligible. This is consistent with previous findings [41], where finetuning self-supervised models with 100% labeled samples offered little improvement in downstream tasks.

**Automatic weighted loss improves finetuning performance**. Pretraining with only contrastive learning resulted in $0.868 \pm 0.013$ AUC and $0.482 \pm 0.032$ AP. Adding context restoration improved AUC by 0.5% and AP by 2.6%. Adding rotation prediction further improved AUC by 0.6% AP by 2.2%. Finally, we compare finetuning our model using proposed automatic weighted loss and equally weighting each pretext loss. We find that our proposed loss improves model performance in 1.7% gain in AUC and 1.6% gain in AP.

**Table.2.** Data efficiency of self-supervised learning in csPCa detection under different training settings. RandInit, CL, CR, Rot, AWL are short for random initialization, contrastive learning, context restoration, rotation and our proposed automatic weighted loss.

| Train | Initialization | AUC | AP |
|---|---|---|---|
| 25% | RandInit. | $0.771 \pm 0.018$ | $0.369 \pm 0.064$ |
|  | CL+CR+Rot with AWL | $0.818 \pm 0.014$ | $0.411 \pm 0.067$ |
| 50%  | RandInit.                    | 0.819 ± 0.022     | 0.431 ± 0.056     |
|      | CL+CR+Rot with AWL           | 0.846 ± 0.015     | 0.465 ± 0.049     |
|      | RandInit.                    | 0.880 ± 0.010     | 0.543 ± 0.042     |
|      | CL+CR+Rot with AWL           | 0.888 ± 0.013     | 0.545 ± 0.060     |
| 100% | CL+CR+Rot with equal weight  | 0.878 ± 0.015     | 0.531 ± 0.057     |
|      | CL+CR                        | 0.872 ± 0.015     | 0.507 ± 0.055     |
|      | CL                           | 0.868 ± 0.013     | 0.482 ± 0.032     |

### 3.2.4 Performance on Prostate158

As described in 3.2.1, we directly evaluated all models trained from PI-CAI dataset on Prostate158 to gauge generalization.

**Patient-based diagnosis:** Self-supervised CSwin UNet reached 0.79 AUC on Prostate158, ahead of Swin UNETR ($p<0.005$), UNet ($p<0.005$), Attention UNet ($p<0.005$), DynUNet ($p<0.005$) (Fig. 8). Self-supervision improved CSwin UNet's performance by 1.8% in AUC ($p<0.005$).

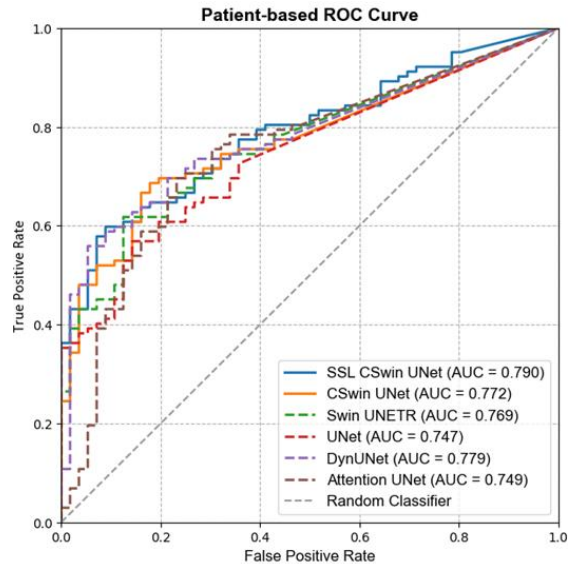

**Figure 8.** Patient-based ROC analysis of csPCa detection on external data using the proposed models and other comparable methods.

**Lesion-based localization:** From PR analysis, self-supervised CSwin UNet reached 0.451 AP on Prostate158, ahead of Swin UNETR ($p<0.005$), UNet ($p<0.005$), Attention UNet ($p<0.005$), DynUNet ($p<0.005$) (Fig. 9). Without self-supervision, CSwin UENTR reached 0.363 AP ($p<0.005$). Our proposed self-supervision method improved CSwin UNet's AP by 8.8%, leading all other models by 10.1%-24.8%. Visualization of detection maps show that self-supervised pretraining improved detected lesion's overlap with ground truth (Fig.9).

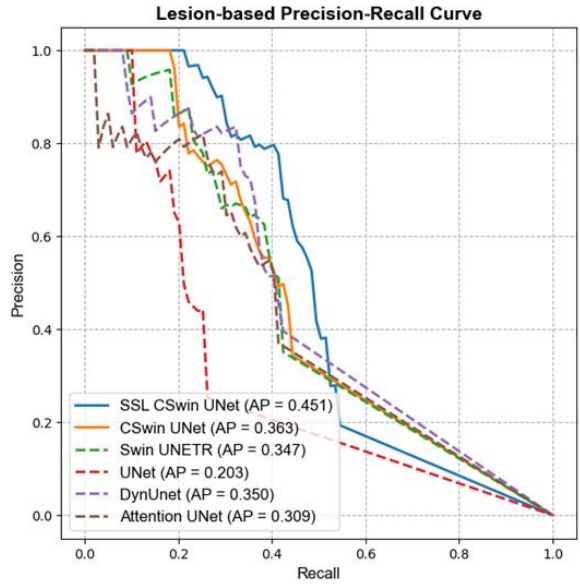

**Figure 9**. Lesion-based PR analysis of csPCa detection on external data using the proposed models and other comparable methods.

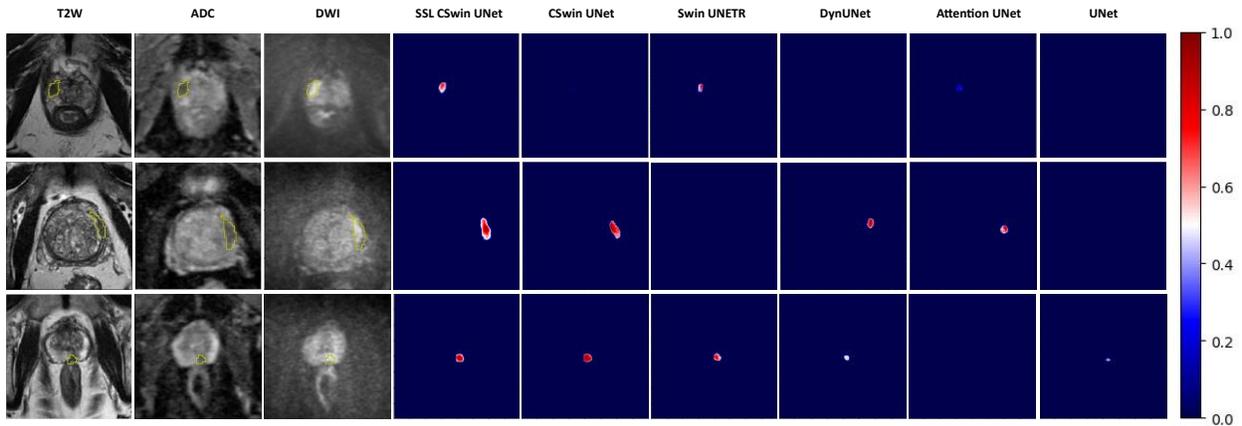

**Figure 10**. Visualization of csPCa detection maps from our proposed systems and comparable methods. GradCAMs and their corresponding T2W, DWI and ADC scans for three patient cases from Prostate158 are shown above. CSwin UNet detected csPCa while other candidate models failed (*middle row, bottom row*). Self-supervised pretraining improved lesion detection overlap with ground truth (*middle row*).

## 4. Discussion

Recent success of vision transformers has inspired a number of transformer-based methods for medical image analysis [42-44]. However, attention-based CNN models still dominate the field of csPCa detection or classification. Compared to CNNs, transformers utilize a sequence-to-sequence approach to effectively capture global context and encode spatial information between patches which are important for image segmentation. Our work first presents a novel cross-shaped window transformer architecture for csPCa detection. Compared to Swin transformer, CSwin benefits from a larger receptive field due to calculating self-attention in parallel in longitudinal, vertical and horizontal stripe, thereby improving network performance. Also, we propose to integrate cosine attention to CSwin transformer which further improves detection performance.

In the medical domain, procurement of expert-annotated data remains a challenge for developing robust AI systems. Self-supervised learning has emerged as a promising approach to effectively training on unlabeled data. However, previous work focuses primarily on pretraining with only one pretext task. Tang et al. [20] combined three pretext tasks using hyperparameter search, which can be practically difficult due to limited computational resources. We present a novel approach to combine pretext tasks using the concept of auxiliary tasks to improve representation learning. Inspired by multi-task learning [28], we formulate an automatic weighted loss function using uncertainty estimation. This approach unifies the various pretext tasks and learns the optimal weights for each task for pretraining. During evaluation, self-supervised models demonstrate good sample efficiency on limited data and outperform baselines with random initializations. SSL pretrained models also generalize well when evaluated on external data.

Our method still has some limitations. We note that the gains in performance from SSL pretraining gradually decrease when finetuning on more labelled samples. This observation is consistent with previous findings in [41], where the pretrained dataset are in-distribution with fine-tuning dataset. When evaluating PI-CAI trained models on Prostate158, we find an overall decrease in performance. We primarily attribute the causes to the disparity between the histologically-confirmed training/validation annotations in PI-CAI and the radiologically-estimated testing annotations in Prostate158 (see Section 3.1.1 and 3.1.2). Other factors including different MRI-scanners and imaging protocols could also affect the outcome of our model inference. SSL requires a large unlabeled dataset that covers a wide range of data distribution and diversity to fully leverage its efficacy. In future work, we aim to pretrain our network with several multi-institutional prostate MRI data to further explore this technique.

## 5. Conclusion

In this paper, we propose a novel transformer-based architecture CSwin UNet for end-to-end csPCa detection and demonstrate the effectiveness of self-supervised pretraining in improving model generalizability. We propose to use automatic weighted loss to dynamically adjust weights of SSL pretext tasks during training to improve representation learning. We evaluate our method on two multi-institutional public datasets and show that it outperforms state-of-the-art methods in detection metrics. We show our self-supervised pretraining method can achieve better generalization to external data than existing methods.

## Acknowledgments

This research is supported in part by the National Institutes of Health under Award Number R01CA215718, R56EB033332 and R01EB032680.